\documentclass[aps,preprintnumbers,epsfig,prl,twocolumn,showpacs]{revtex4}
\usepackage{amssymb}
\usepackage{graphics}
\usepackage{graphicx}
\usepackage{epsfig}

\begin{document}

\title{Dynamical analysis of the nonlinear growth of the $m=n=1$ resistive internal mode}

\author{M.-C. Firpo}
\altaffiliation[Present address: ]{LPGP, B\^{a}t. 210, UPS, F-91405 Orsay}
\email{marie-christine.firpo@lpgp.u-psud.fr}
\author{B. Coppi}
\affiliation{Massachusetts Institute of Technology, Cambridge, MA 02139-4307}
\date{\today}

\begin{abstract}
A dynamical analysis is presented that self-consistently takes into account the
motion of the critical layer, in which the magnetic field reconnects, to
describe how the $m=n=1$ resistive internal kink mode develops in the nonlinear
regime. The amplitude threshold marking the onset of strong nonlinearities due
to a balance between convective and mode coupling terms is identified. We
predict quantitatively the early nonlinear growth rate of the $m=n=1$ mode
below this threshold.
\end{abstract}



\maketitle

The large scale dynamics and confinement properties of tokamak plasmas
depend intimately on the behavior of $m=n=1$ magnetohydrodynamic (MHD)
internal kink modes. This has motivated an intense, long-lasting,
experimental and theoretical research, notably devoted to study their
implication in magnetic reconnection or as triggers of the sawtooth
oscillations and crashes. These phenomena typically proceed beyond the
linear regime, that is now rather well understood but assumes very small
amplitudes of the modes. To offer a quantitative, predictive description of
their nonlinear manifestations remains a difficult objective of both
academic interest and very practical importance. This is especially relevant
for the design of fusion burn experiments in which the fulfilment of linear
stability constraints is challenged by the search for ignition. Such devices
are thus expected to operate at best close to marginal stability for the $%
m=n=1$ ideal mode so that nonlinear effects come into play for fairly small
values of the mode amplitude \cite{Coppi02,Odblom02}.

In this Letter, we focus on the $m=n=1$ resistive mode \cite{Coppi76} in which
a finite resistivity $\eta $ destabilizes the otherwise marginally stable ideal
MHD internal kink mode. Since Kadomtsev's scenario \cite{Kadomtsev} predicting
the complete reconnection of the helical flux within the $q=1$ surface on a
timescale of order $\eta ^{-1/2}$, that later appeared too large to account for
observations, the nonlinear behavior of the $m=n=1$ mode has become a somewhat
controversial issue. Some numerical simulations suggested that the mode still
grows exponentially into the nonlinear regime \cite{waddell} which was
supported by a theoretical model \cite{Hazeltine86}. Later some analytic
studies \cite{Waelbroeck89}, supported by numerical simulations
\cite{Biskamp91}, rather predicted a transition to an algebraic growth early in
the nonlinear stage. This result was challenged by Aydemir's recent simulations
using a dynamical mesh \cite{Aydemir97}. These did show the linear exponential
stage evolving towards an algebraic stage, yet this was brutally interrupted by
a second nonlinear exponential growth. A modified Sweet-Parker model was able
to fit continuously both stages of evolution \cite{Aydemir97} and the
transition related to a change in the geometry of the current sheet
\cite{Wang99}. However, some fundamental questions remain unanswered or
unclear. Among them, how to relate the transition threshold with $\eta $ ? or
what is the role of the $q$-profile ? The aim of this Letter is to describe
analytically how the $m=n=1$ resistive mode develops in the nonlinear regime,
by focusing on the equations controlling plasma dynamics.

We consider the low-$\beta $ reduced MHD equations
\begin{eqnarray}
\frac{\partial U}{\partial t} &=&\left[ \phi ,U\right] +\left[ J,\psi \right]
\label{finsys_1} \\
\frac{\partial \psi }{\partial t} &=&\left[ \phi ,\psi \right] +\eta
(J-J_{0})  \label{finsys_2}
\end{eqnarray}
assuming helical symmetry \cite{FirpoMIT}. Only a single angular variable is
then involved in the problem, namely the helical angle $\alpha \equiv
\varphi -\theta $, with $\varphi $ the toroidal and $\theta $ the poloidal
angles. $U=\nabla _{\bot }^{2}\phi $ is the vorticity and $J=\nabla _{\bot
}^{2}\psi $ the helical current density, with $\nabla _{\bot }^{2}\equiv
r^{-1}\partial _{r}r\partial _{r}+r^{-2}\partial _{\alpha }^{2}$. Time is
normalized by the poloidal Alfv\'{e}n time ($t\rightarrow t/\tau _{Hp}$),
the radial variable by the minor radius ($r\rightarrow r/a$) and $\eta $ is
the dimensionless resistivity, inverse of the magnetic Reynolds number $S$ ($%
\eta \equiv S^{-1}=\tau _{Hp}/\tau _{R}$) with the poloidal Alfv\'{e}n time $%
\tau _{Hp}=\left( \mu _{0}\rho _{0}\right) ^{1/2}R/B_{0\varphi }$ and
resistive time $\tau _{R}=\mu _{0}a^{2}/\eta _{0}$. The Poisson brackets are
defined by $\left[ \phi ,U\right] =-\mathbf{\hat{\varphi}}\cdot \left(
\mathbf{\nabla }_{\bot }\phi \times \mathbf{\nabla }_{\bot }U\right)
=r^{-1}\left( \partial _{r}\phi \partial _{\alpha }U-\partial _{r}U\partial
_{\alpha }\phi \right) $. $\phi $ and $\psi $ are the plasma velocity and
helical magnetic field potentials expressed in cylindrical coordinates, so
that the velocity is $\mathbf{v}=\mathbf{\hat{\varphi}}\times \mathbf{\nabla
}_{\bot }\phi $ and the magnetic field is $\mathbf{B}=B_{0\varphi }\mathbf{%
\hat{\varphi}}+\mathbf{\hat{\varphi}}\times \mathbf{\nabla }_{\bot }\left(
\psi -r^{2}/2\right) $.

We consider MHD equilibria given by $\phi _{0}=0$ and by an helical magnetic
flux $\psi _{0}\left( r\right) $, related to the safety profile $q(r)$
through $d_{r}\psi _{0}=r\left[ 1-1/q(r)\right] $, such that $q=1$ for an
internal radius $r=r_{s0}$. Thus $d_{r}\psi _{0}\left( r_{s0}\right) =0$.
This means that the low-frequency ideal linear equations associated to (\ref
{finsys_1})-(\ref{finsys_2}) are singular at $r=r_{s0}$, with a formally
diverging current density. This marks the presence of a critical layer in
which the dynamics differs considerably from the outer one and where
resistivity enters to cure the singularity.

We wish to analyse perturbatively the time evolution of the $m=1$ mode. For
this, we assume that only the $m=1$ mode is destabilized initially with an
amplitude $A_{0}$, neglect all ideal MHD transients and restrict to the
linear resistive timescale $\tau \equiv \eta ^{1/3}t$. We do not consider
the somehow ill-posed, singular limit $\eta \rightarrow 0$, but instead
realize that \textit{two} small parameters are indeed competing in this
problem, namely the small \textit{given} resistivity $\eta $ and the
time-dependent amplitude $A(\tau )$ of the linear $m=1$ mode. This
introduces some subtleties in the amplitude expansion. The order one
solution is given by linear theory using an asymptotic analysis \cite
{Coppi76} to match inner and outer solutions. Excitation of the $m=1$ mode
leads to a self-consistent correction to the location of the critical layer.
One estimates the amplitude threshold, scaling with $\eta $, at which next
order solution is required and the procedure iterated. Separability in time
and space propagates at each order resulting in an amplitude expansion in $A$%
. As in any perturbative approach, the solution is formally known when the
order one solution is. This is given by the linear theory reviewed now.

Let $f_{n}^{(m)}$ be the projection on $\exp (im\alpha )$ of any function $f$
at order $n$. In the inner resistive layer, Eqs. (\ref{finsys_1})-(\ref
{finsys_2}) read
\begin{eqnarray}
\left[ \frac{\partial }{\partial \tau }\frac{\partial ^{2}}{\partial x^{2}}%
\phi _{1}^{(1)}+i\kappa _{0}x\frac{\partial ^{2}}{\partial x^{2}}\psi
_{1}^{(1)}\right] w^{-1} &=&0  \label{lin_1} \\
\left[ \frac{\partial }{\partial \tau }\psi _{1}^{(1)}+i\kappa _{0}x\phi
_{1}^{(1)}-\frac{\partial ^{2}}{\partial x^{2}}\psi _{1}^{(1)}\right] w &=&0
\label{lin_2}
\end{eqnarray}
where we define $\kappa _{0}\equiv \psi _{0}^{\prime \prime }\left(
r_{s0}\right) /r_{s0}$. In these equations, $x$ is the stretched coordinate $%
x=\left( r-r_{s0}\right) /w$ and $w\equiv \eta ^{1/3}$ the magnitude of the
width of the critical layer giving the maximal resistive ordering \cite
{Coppi76} in (\ref{lin_1})-(\ref{lin_2}). In the layer, radial derivatives
are large, since $\partial _{r}=w^{-1}\partial _{x}$ and (\ref{lin_1})-(\ref
{lin_2}) are the dominant equations for $w\ll 1$. There is one unstable
solution, the $m=1$ resistive mode, with growth rate $\hat{\gamma}%
_{L}=\kappa _{0}^{2/3}=q^{\prime }(r_{s0})^{2/3}$. Real space potentials
read
\begin{eqnarray}
\psi _{1}\left( x,\alpha ,\tau \right) &=&A_{0}\exp \left( \hat{\gamma}%
_{L}\tau \right) g_{L}\left( \frac{\kappa _{0}^{1/3}x}{\sqrt{2}}\right) \cos
\alpha  \label{psi-1rst-order} \\
\phi _{1}\left( x,\alpha ,\tau \right) &=&-\frac{A_{0}}{\sqrt{2}}\exp \left(
\hat{\gamma}_{L}\tau \right) g_{L}^{\prime }\left( \frac{\kappa _{0}^{1/3}x}{%
\sqrt{2}}\right) \sin \alpha  \label{phi-1rst-order}
\end{eqnarray}
where $g_{L}$ is the function
\begin{equation}
g_{L}\left( s\right) =\frac{s}{2}\mathop{\mathrm{erfc}}\left( s\right) -%
\frac{1}{2\sqrt{\pi }}\exp (-s^{2}).  \label{def-gL}
\end{equation}
This solution was chosen to satisfy the matching asymptotic conditions $%
g_{L}^{\prime }\left( -\infty \right) =1$ and $g_{L}^{\prime }\left( +\infty
\right) =0$. This analysis has to be complemented with the derivation of the
outer solution. On the resistive timescale, this amounts to solve, at
leading (zero) order in $w$, a linear system of ideal MHD equilibria,
singular at $r=r_{s0}$ \cite{FirpoMIT}. This illustrates the passive
character of the outer domain. We only retain here that, given the
asymptotic and boundary conditions imposing $\psi _{1}^{\prime
(1)}(r_{s0}^{+})=0$ and $\psi _{1}^{(1)}(1)=0$, the outer linear $m=1$
solution $\psi _{1}^{(1)}(r)$ is identically vanishing for $r_{s0}<r\leq 1$.

Linear theory breaks down when, in the resistive critical layer, nonlinear
terms due to mode couplings, e.g. in Eq. (\ref{finsys_1}) $\left[ \phi
_{1},U_{1}\right] \sim w^{-3}A^{2}$, balance linear ones, i.e. $\kappa
_{0}xw\partial _{\alpha }J_{1}\sim A/w$ in Eq. (\ref{lin_1}). Thus $A(\tau )=%
\mathcal{O}(\eta ^{2/3})$ marks the onset of second order terms. Before
pursuing the analysis on the critical layer, we need to track it and
self-consistently estimate its location. The total magnetic flux in the
critical layer is now $\psi \left( x,\alpha ,\tau \right) =\eta ^{2/3}\psi
_{0}^{\prime \prime }\left( r_{s0}\right) x^{2}/2+\psi _{1}\left( x,\alpha
,\tau \right) $. To follow continuously the linear stage, we define the
'backbone' $r_{s}\left( \alpha ,\tau \right) $ of the critical layer as the
'neutral' field line with $\partial _{r}\psi \left( r_{s}\right) \equiv 0$.
Writing $r_{s1}\left( \alpha ,\tau \right) =r_{s}\left( \alpha ,\tau \right)
-r_{s0}=wx_{1}(\alpha ,\tau )$ with $\partial _{x}\psi (x_{1})=0$, this
gives
\begin{equation}
r_{s1}\left( \alpha ,\tau \right) \simeq -\frac{A\left( \tau \right) }{\eta
^{1/3}}\frac{\kappa _{0}^{1/3}g_{L}^{\prime }\left( 0\right) }{\sqrt{2}\psi
_{0}^{\prime \prime }\left( r_{s0}\right) }\cos \alpha  \label{rs_1rst_order}
\end{equation}
which relates to the shift of the core plasma inside the $q=1$ surface due
to the kink instability. Then the x-point shift $r_{s1}\left( \alpha =\pi
,\tau \right) $ goes like $A(\tau )/\eta ^{1/3}$, consistently with
Aydemir's numerical results \cite{Aydemir97}. Thus the critical radius
starts to leave the linear critical layer band, centered on $r_{s0}$, when $%
r_{s1}\left( \alpha ,\tau \right) $ becomes of the order $\eta ^{1/3}$ for
some $\alpha $, that is when $A\left( \tau \right) \gtrsim \eta ^{2/3}$.
This is again the threshold marking the end of the linear stage. We need now
to define a generalized stretched coordinate in the critical layer as $%
x=\left( r-r_{s}\left( \alpha ,\tau \right) \right) /w$. The replacements $%
\partial _{\tau }\rightarrow \partial _{\tau }-w^{-1}\left( \partial
r_{s}/\partial \tau \right) \partial _{x}$ and $\partial _{\alpha
}\rightarrow \partial _{\alpha }-w^{-1}\left( \partial r_{s}/\partial \alpha
\right) \partial _{x}$ are then required \cite{note1}.

The second order critical layer equations involve an inhomogeneous part
composed of quadratic terms in the order one solutions (\ref{psi-1rst-order}%
), (\ref{phi-1rst-order}) and (\ref{rs_1rst_order}). This acts to force the
growth of the $m=0$ and $m=2$ perturbations but brings no contribution to
the $m=1$ dynamics. Therefore the $m=1$ equations (\ref{lin_1})-(\ref{lin_2}%
) are unchanged, except that, due to the motion of the critical layer (\ref
{rs_1rst_order}), one needs to replace $\kappa _{0}$ in (\ref{lin_1})-(\ref
{lin_2}) by the time-dependent average
\begin{equation}
\kappa ^{(0)}(\tau )\equiv \frac{1}{2\pi }\int\limits_{0}^{2\pi }\frac{%
\partial _{r}^{2}\psi \left[ r_{s}\left( \alpha ,\tau \right) \right] }{%
r_{s}\left( \alpha ,\tau \right) }d\alpha .  \label{def-kappa-t}
\end{equation}
This introduces a generalized linear system of equations. Neglecting the
initially zero amplitudes of the $m=0$ and $m=2$ perturbations in front of $%
A(\tau )$, the second order correction to the location of the critical layer
is given by $r_{s2}\left( \alpha ,\tau \right) \simeq -\left( 2\psi
_{0}^{\prime \prime }\left( r_{s0}\right) \right) ^{-1}\psi _{0}^{\prime
\prime \prime }\left( r_{s0}\right) r_{s1}\left( \alpha ,\tau \right) ^{2}$.
The validity threshold of the second order solution is reached when the
instantaneous critical line moves out of the critical layer of width $w$
centered on $r_{s0}+r_{s1}\left( \alpha ,\tau \right) $ for some $\alpha $.
This corresponds to $r_{s2}\left( \alpha ,\tau \right) \sim w$, that is to $%
r_{s1}\left( \alpha ,\tau \right) ^{2}\sim \eta ^{1/3}$, which gives $A(\tau
)=\mathcal{O}\left( \eta ^{1/2}\right) $. This threshold in the amplitude of
the linear $m=1$ mode marks the onset of third order terms, that will
contribute again to the $m=1$ dynamics. Its brutal manifestation is visible
on Aydemir's plots \cite{Aydemir97}. They clearly report a transition in the
$m=1$ kinetic energy when this becomes of order $\eta /2$ \cite{note2},
namely around $5\times 10^{-8}$ for $\eta =10^{-7}$ and around $5\times
10^{-6}$ for $\eta =10^{-5}$.
\begin{figure}[tbp]
\centering
\includegraphics[width=8cm]{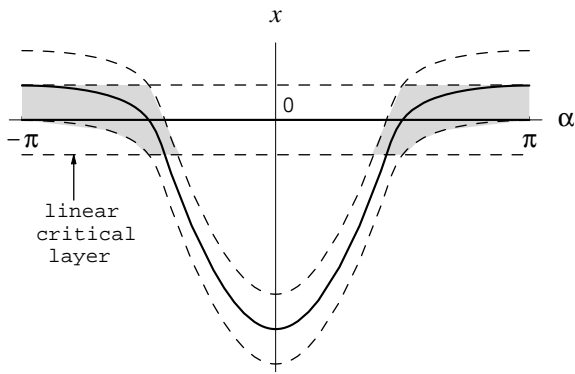}
\caption{Picture in the $(x,\protect\alpha )$ space of the initial linear
critical layer and of a nonlinear one centered on the instantaneous
transverse neutral field line (in bold). The grey region represents their
overlapping domain within which the gradients of linear potentials are $%
\mathcal{O}(w^{-1})$-large.}
\label{fig-criti-layer}
\end{figure}

At third order, cubic terms in the order one solutions or quadratic terms
coupling the $m=0$ and $m=2$ second order terms to the $m=1$ first order
ones appear in the inhomogeneous part of the critical layer equations and
modify the $m=1$ dynamics. These terms involve some radial derivatives, e.g.
$\partial _{r}\phi _{1}^{(1)}$, that are $\mathcal{O}(w^{-1})$-large only
within the linear layer. Locality enters here the analysis since the
dominant contribution of these mode coupling terms comes from the localized
zone in $(r,\alpha )$ where the instantaneous and linear critical layers
overlap. This is depicted by the grey shaded region in Fig. \ref
{fig-criti-layer}. The novelty is that, in this region, mode couplings are
now able to balance convective derivatives, both being dominant with respect
to linear terms. More explicitly, while, e.g. in the Eq. (\ref{finsys_1})
written in the region where the instantaneous and linear critical layers
overlap, the magnitude of linear terms is $\partial _{\tau }\partial
_{r}^{2}\phi ^{(1)}\sim w^{-2}A(\tau )$, convective terms are of the order
of $\partial _{\tau }r_{s2}^{(0)}\partial _{r}^{3}\phi _{1}^{(1)}\sim
w^{-5}A^{3}$. Thus linear terms become negligible for $A(\tau )\gg \eta
^{1/2}$, which marks the onset of the fully nonlinear regime for the $m=1$
mode. Moreover, convective terms, e.g. $\partial _{\tau
}r_{s2}^{(0)}\partial _{r}^{3}\phi _{1}^{(1)}\sim w^{-5}A^{2}\partial _{\tau
}A$, equilibrate mode coupling terms, such as $-r_{s0}^{-1}\partial _{r}\phi
_{1}^{(-1)}\partial _{\alpha }r_{s1}^{(1)}\partial _{r}U_{1}^{(1)}\sim
w^{-5}A^{3}$ coming from $\left[ \phi ,U\right] $ in the shear-Alfv\'{e}n
law (\ref{finsys_1}). The nonlinear growth rate derives from this balance.
As $\kappa ^{(0)}(\tau )$ is no longer involved in those convective and mode
coupling terms, there is no extra time-dependence in the dominant equations,
so that the nonlinear growth rate is just equal, by continuity, to the
growth rate of the $m=1$ mode when $A(\tau )$ becomes of order $\eta ^{1/2}$%
. Its value depends notably on the equilibrium $q$-profile as we shall see
below. After some spatial averaging, a rough summary of the time evolution
of the $m=1$ mode amplitude may be then finally written as
\begin{equation}
\frac{dA}{dt}-\gamma \left( t\right) A+\frac{c}{\eta }A^{2}\left( \frac{dA}{%
dt}-\gamma \left( t_{NL}\right) A\right) =0,  \label{balanceNL}
\end{equation}
where the initial value of the growth rate $\gamma \left( 0\right) $ is $%
\gamma _{L}$ and where the early time dependence of $\gamma $ comes from the
motion of the critical layer and is computed quantitatively below. In Eq. (%
\ref{balanceNL}), $c$ is a constant of order one and $t_{NL}$ denotes the
(magnitude of the) time at which $A$ becomes of order $\eta ^{1/2}$. Eq. (%
\ref{balanceNL}) describes effectively the transition between two (almost)
exponential stages. Because $\phi _{3}^{(1)}$and $\psi _{3}^{(1)}$ are zero
at the onset of the third order regime, Eq. (\ref{balanceNL}) remains valid
during some stage even if the structure and scaling of the critical layer
should substantially change as the generalized linear stage is left.

For the convective exponential stage to be fully valid, the overlap between
the linear and instantaneous critical layers should be large enough. One
expects then a qualitatively different late behavior of the $m=1$ dynamics
if the x-point region is far away from the linear layer when $A(\tau )=%
\mathcal{O}\left( \eta ^{1/2}\right) $, that is, due to (\ref{rs_1rst_order}%
), if $\eta ^{-1/6}\ggg 1$. This regime is extremely challenging to reach
numerically but may be satisfied in tokamak plasmas.

We finally examine the early nonlinear effects on the growth rate of the $%
m=1 $ mode due to the motion of the critical layer. This amounts to solve
the system of differential equations (\ref{lin_1})-(\ref{lin_2}) for $\kappa
_{0} $ replaced with $\kappa ^{(0)}\left( \tau \right) $, defined in (\ref
{def-kappa-t}). It can be checked that, as long as the order of magnitude of
$A\left( \tau \right) $ is lower than $\eta ^{1/2}$, $\kappa ^{(0)}\left(
\tau \right) $ may be approximated by $\left( 2\pi \right)
^{-1}\int\nolimits_{0}^{2\pi }r_{s}\left( \alpha ,\tau \right) ^{-1}\psi
_{0}^{\prime \prime }\left[ r_{s}\left( \alpha ,\tau \right) \right] d\alpha
$ at leading order. This expression will be retained in the numerical
computations. The time-dependent growth rate is defined as $\hat{\gamma}%
\left( \tau \right) \equiv d_{\tau }A/A.$ In this generalized linear system,
there is one condition shared with the linear derivation: for a solution in
separate variables $\tau $ and $x$, it is that $\hat{\gamma}\left( \tau
\right) /\kappa \left( \tau \right) $ be constant. This constant is then
fixed by continuity with the linear solution at time zero giving
\begin{equation}
\frac{\hat{\gamma}\left( \tau \right) }{\kappa ^{(0)}\left( \tau \right) }=%
\frac{\hat{\gamma}_{L}}{\kappa _{0}}=\kappa _{0}^{-1/3}.  \label{relation-1}
\end{equation}
Here one implicitly assumes that the spatial part of the linear
eigenfunctions remains valid \cite{note3}. The instantaneous critical radius
is $r_{s}\left( \alpha ,\tau \right) =r_{s0}+\eta ^{1/3}x_{s}\left( \alpha
,\tau \right) $ where $x_{s}(\alpha ,\tau )$ is given by the approximate
expression
\begin{equation}
x_{s}\left( \alpha ,\tau \right) =H^{-1}\left( -\frac{A(\tau )\kappa
_{0}^{1/3}\cos \alpha }{\eta ^{2/3}\sqrt{2}\psi _{0}^{\prime \prime }(r_{s0})%
}\right) .  \label{expres-xs}
\end{equation}
$H^{-1}$ denotes the inverse of the monotonously growing function defined by
$H(x)\equiv x/g_{L}^{\prime }\left( \kappa _{0}^{1/3}x/\sqrt{2}\right) $. Due
to the asymmetric nature of the $m=1$ resistive eigenfunctions (\ref {def-gL}),
$H^{-1}(x)$ is very asymmetric, grossly equal to $x$ below $x=0$ and
exponentially small above. This confers a much more important weight on
negative arguments of $H^{-1}$ than on positive ones in the averaging (\ref
{def-kappa-t}). The magnetic island has thus a higher effective contribution to
the early nonlinear correction of the growth rate than the region of
x-point. A rough estimate of the angular average of $x_{s} $ is given by $%
x_{s}^{(0)}(\tau )\simeq -\left( 2\pi \right) ^{-1}A\left( \tau \right)
\kappa _{0}^{1/3}/\left( \eta ^{2/3}\sqrt{2}\psi _{0}^{\prime \prime
}(r_{s0})\right) \int\nolimits_{-\pi /2}^{\pi /2}d\alpha \cos \alpha $. Eq. (%
\ref{relation-1}) defines a first order differential equation in $A(\tau )$
that admits then the approximate form $\hat{\gamma}\left( \tau \right)
\simeq \hat{\gamma}_{L}+\eta ^{1/3}d_{r}\left[ r^{-1}\psi _{0}^{\prime
\prime }(r)\right] \left( r_{s0}\right) x_{s}^{(0)}(\tau )$. Going back to
time $t$ and to $\gamma _{L}\equiv \eta ^{1/3}\hat{\gamma}_{L}$, this gives
\begin{equation}
\frac{dA}{dt}\simeq \gamma _{L}A(t)-C_{0}A(t)^{2}  \label{relat_2}
\end{equation}
where $C_{0}=$ $\left( q_{0}^{\prime }+r_{s0}q_{0}^{\prime \prime
}-2r_{s0}q_{0}^{\prime 2}\right) /\left( \pi \sqrt{2}r_{s0}^{2}q_{0}^{\prime
2/3}\right) $ and the index $0$ denotes an evaluation at $r_{s0}$. Eq. (\ref
{relat_2}) shows the first nonlinear contribution to the $m=1$ evolution.
The early behavior of the $m=1$ growth rate is thus $\gamma (t)\simeq \gamma
_{L}-C_{0}A_{0}\exp \left( \gamma _{L}t\right)$. In order to check
numerically these analytic predictions for the generalized linear stage,
that brings the first nonlinear contributions to the growth rate, we used
Aydemir's initial conditions \cite{Aydemir97}. The safety profile is $%
q(r)=q_{m}\left\{ 1+r^{4}\left[ \left( q_{a}/q_{m}\right) ^{2}-1\right]
\right\} ^{1/2}$ with $q_{m}=0.9$, $q_{a}=3$, giving $C_{0}>0$. The
differential equation (\ref{relation-1}) was integrated numerically for
$A_{0}=\sqrt{2}\times 10^{-5.5}$ corresponding to an initial kinetic energy in
the $m=1$ mode of the order $10^{-11}$.
\begin{figure}[tbp]
\centering
\includegraphics[width=8cm]{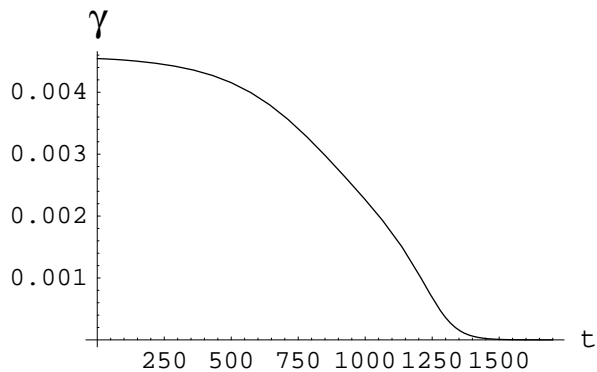}
\caption{Analytic nonlinear growth rate corresponding to the initial conditions
used in Ref. \protect\cite{Aydemir97} and resistivity $\protect\eta =10^{-7}$,
neglecting third order convective effects coming into play when $A(t)$ becomes
of order $\protect\eta ^{1/2}$. This occurs for $t\simeq 1000$.}
\label{gamma_NL_plot}
\end{figure}
The nonlinear growth rate $\gamma \left( t\right) \equiv \eta ^{1/3}\hat{%
\gamma}(\tau )$ is plotted on Fig. \ref{gamma_NL_plot} for $S=10^{7}$. This
curve appears to be in fine agreement with the Figure 1 of Ref. \cite
{Aydemir97} for times $t$ roughly below 1000 Alfv\`{e}n times.

\begin{figure}[h]
\centering
\includegraphics[width=8cm]{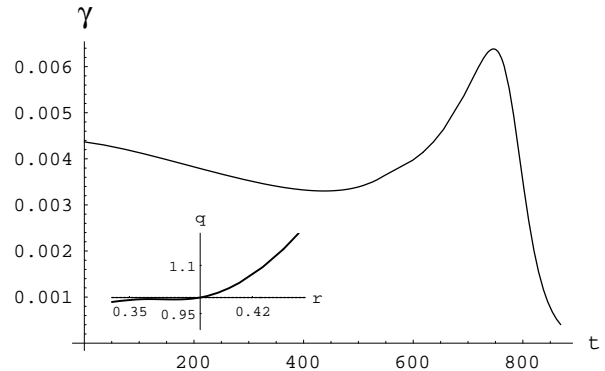}
\caption{Analytic nonlinear growth rate for the same initial values as in
Fig. \ref{gamma_NL_plot} but with a modified equilibrium safety profile $%
q(r) $. Its behavior around $r_{s0}$ is plotted in the insert.}
\label{gamma_NL_modif}
\end{figure}

Fig. \ref{gamma_NL_modif} illustrates the influence of the $q$-profile around
$r_{s0}$ on the time evolution of $\gamma$ due to (\ref{def-kappa-t}). A sudden
bump in the nonlinear growth could thus even be observed, before the onset of
convective effects, for the special shape of $q$ chosen in Fig.
\ref{gamma_NL_modif}. Moreover, some $q$-profile may induce a saturation of $%
A$ below the convective threshold and lead to partial reconnection. Most
importantly, the approach described here may be transposed to model the
early nonlinear behavior of a variety of internal kinks such as two-fluid
\cite{Aydemir92,Rogers96,Biskamp97} and/or collisionless \cite{Cafaro98}
models.

\vskip 1.5cm

Discussions with L. Sugiyama are gratefully acknowledged. MCF thanks A.
Aydemir for several communications on his simulations. This work was
supported in part by the U.S. Department of Energy.


\begin{thebibliography}{99}
\bibitem{note1}  It can be checked that angular contributions in Laplacians
are negligible in the critical layer for $A(\tau )\ll \eta ^{1/3}$.
\bibitem{note3}  This is partly justified by the matching to the vanishing
order one outer solution for $r\geq r_{s0}$.
\end{thebibliography}
\end{document}